\pdfoutput=1

\documentclass[preprint,12pt, a4paper]{elsarticle}

\usepackage{amssymb}
\usepackage{hyperref}
\usepackage{bm}
\usepackage{placeins}
\usepackage{caption}
\usepackage{subfig}
\usepackage{dirtytalk}
\newcommand{\MATLAB}{\textsc{Matlab}}

\journal{SoftwareX}

\begin{document}

\begin{frontmatter}



\title{OpenIPDM: A Probabilistic Framework for Estimating the Deterioration and Effect of Interventions on Bridges}


\author[1]{Zachary Hamida\corref{cor1}}
\author[1]{Blanche Laurent}
\author[1]{James-A. Goulet}
\address[1]{Department of Civil, Geological and Mining Engineering, Polytechnique Montreal, Montreal, Canada}
\cortext[cor1]{corresponding author: zachary.hamida@polymtl.ca}
\begin{abstract}
This paper describes OpenIPDM software for modelling the deterioration process of infrastructures using network-scale visual inspection data. In addition to the deterioration state estimates, OpenIPDM provides functions for quantifying the effect of interventions, estimating the service life of an intervention, and generating synthetic data for verification purposes. Each of the aforementioned functions are accessible by an interactive graphical user interface. OpenIPDM is designed based on the research work done on a network of bridges in Quebec province, so that the concepts presented in the software have been validated for applications in a real-world context. In addition, this software provides foundations for future developments in the subject area of modelling the deterioration as well as intervention planning.
\end{abstract}


\end{frontmatter}


\noindent
\section{Motivation and significance}\label{S:Motive}
Monitoring the deterioration state of bridges is a prerequisite for planning maintenance activities and many other decision-making processes \cite{Chase:2016aa,Zhang:2017aa,Allah-Bukhsh:2019aa}. Visual inspection is a common techniques for performing health monitoring tasks and providing data on a network-scale \cite{Moore:2001aa, Soetjipto:2017aa, MDDQ}. Nonetheless, interpreting information from visual inspection data is known to be challenging \cite{Eugene2015aa,Campbell:2019aa,Hamida:2020aa}. The challenges arise from technical aspects such as, the subjective nature of the evaluation method \cite{AgdasComparison, Bennetts:2018aa, Hamida:2020aa}; and non-technical aspects such as, navigating and performing analyses on thousands of structural elements. Moreover, missing data and outliers can impose inconstancies or cause instability in any type of analyses performed using the inspection data \cite{Hamida:2021wz}. 
In order to address the aforementioned challenges, the software OpenIPDM has been developed to provide access to state-of-the-art deterioration models, all within an interactive graphical user interface (GUI). The GUI facilitates the navigation through the inspection data as well as performing analyses on structural elements, bridges and network of bridges. OpenIPDM also has the capacity to generate synthetic data, perform verification and validation analyses, and automatically manage missing data and outliers. The deterioration model employed in OpenIPDM is based on the SSM-KR framework, which is a hybrid deterioration model that combines state-space models (SSM) with kernel regression (KR) \cite{Hamida:2020ab}. The SSM-KR enables modelling the deterioration condition and speed using visual inspection data, while taking into account the inspectors uncertainties. In addition, the KR framework in SSM-KR allows exploiting the structural similarities among bridges, which contribute to improving the overall predictive capacity of the deterioration model. 

In the context of modelling network-scale deterioration, there exist a handful of softwares with similar functionalities (e.g., Pontis), most of which are: 1) commercial, and 2) rely on frameworks with limited capacity to model the deterioration of infrastructures (e.g., incapacity to estimate and take into account the inspectors' uncertainty) \cite{Thompson:1993tk, Ellis:2008tq, Hamida:2020aa}. OpenIPDM is completely open-source and encompass the most recent developments in the context modelling deterioration based on visual inspections. The software is part of ongoing research, and has already enabled several contributions in the field of structural health monitoring using visual inspections \cite{Hamida:2020aa, Hamida:2020ab, Hamida:2021aa, Hamida:2021wz}.

Aside from performing deterioration analyses on inspection databases, OpenIPDM can serve as a benchmarking tool for developing and enhancing infrastructures deterioration models. In addition, OpenIPDM can facilitate the development and testing of decision-making frameworks, on a small-scale (element-level) or a network-scale. The software collectively provides insights into the development, design and analyses performed on network-scale visual inspections, derived from real case studies.

\section{Software description}
This section provides a general description for the main architecture and functionalities available in the current version of OpenIPDM. 

\subsection{Software architecture}
\begin{figure*}[hbt!]
\centering
\includegraphics[width=1\textwidth]{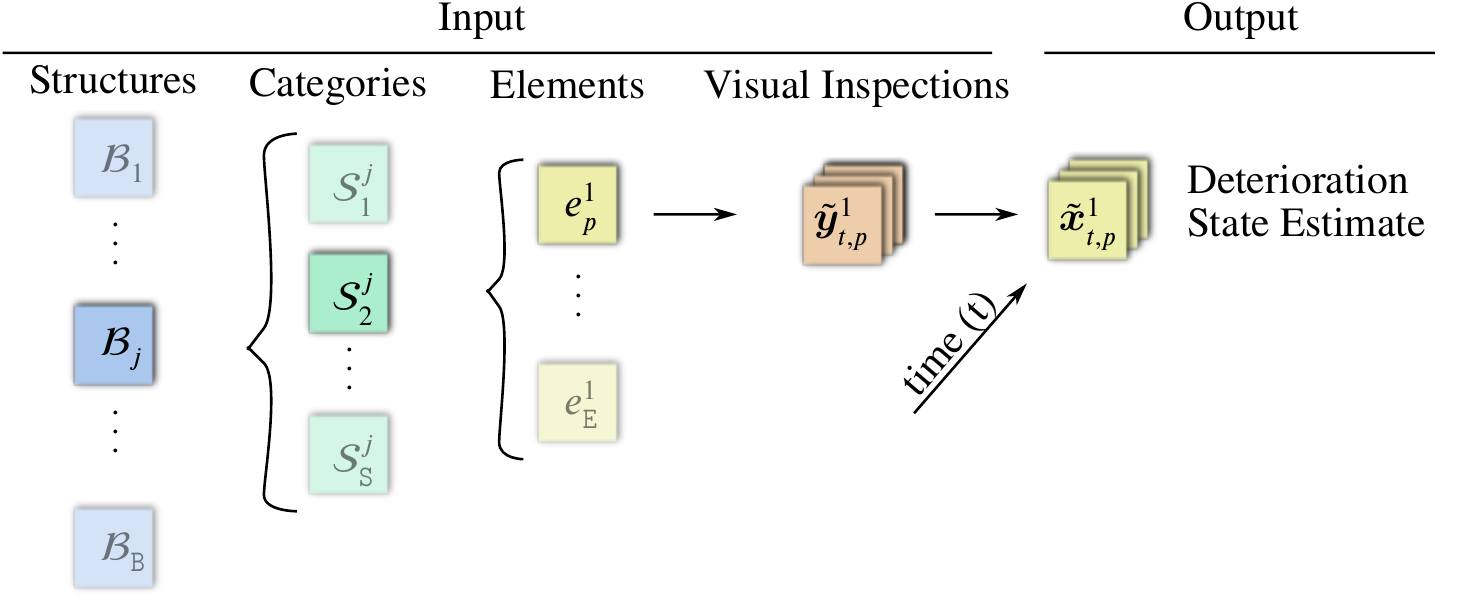}
\caption{Example for the hierarchy of information, starting from bridge $\mathcal{B}_j$, structural category $\mathcal{S}^j_2$, structural element $e_p^j$, and ending with the deterioration state estimates $\tilde{\bm{x}}_{t,p}^j$ over time $t$, based on the inspection data $\tilde{\bm{y}}_{t,p}^j$.}
\label{FIG:data_io}
\end{figure*}

\begin{figure*}[hbt!]
\centering
\includegraphics[width=0.8\textwidth]{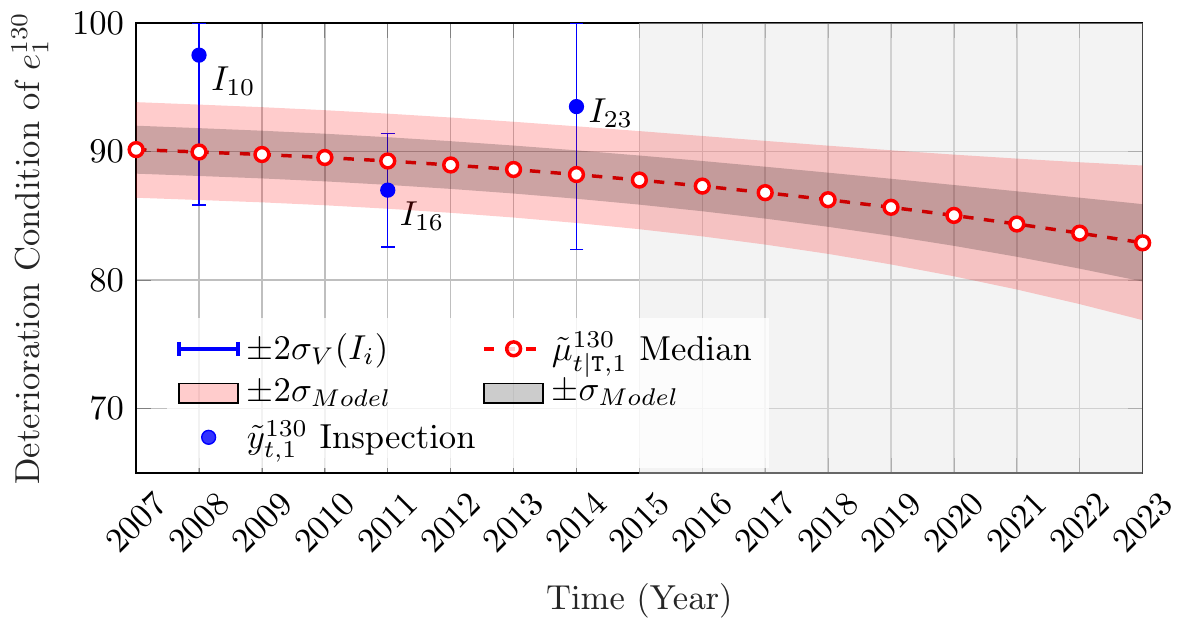}
\caption{Condition deterioration analysis based on observations $\tilde{\bm{y}}_{t,1}^{130}\in[25,100]$ of the structural element $e_1^{130}$ with error bars representing the inspectors estimated uncertainties, and the shaded area representing the forecast period.}
\label{FIG:example_analyses}
\end{figure*}

\begin{figure*}[hbt!]
\centering
\includegraphics[width=1\textwidth]{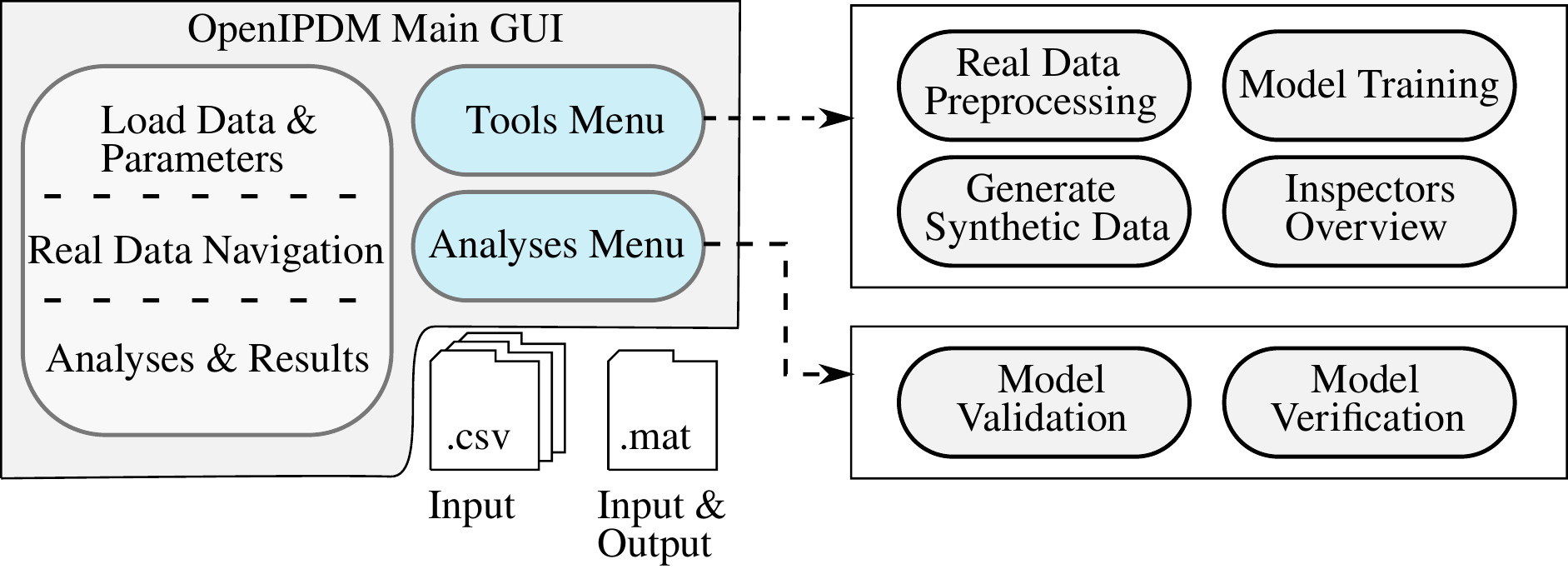}
\caption{OpenIPDM main architecture and functionalities along with the acceptable input-output file formats.}
\label{FIG:main_ui}
\end{figure*}

OpenIPDM is built and compiled using the App Designer in \MATLAB{}, with a complete GUI, providing users an easy access to the different functionalities in the software. The workflows in OpenIPDM operate on two types of data: real data and synthetic data. The real data is a reference for the inspection data collected from real infrastructures, while the synthetic data refers to datasets generated to emulates the real data, with known parameters and access to the true deterioration states \cite{Hamida:2020aa, Hamida:2020ab, Hamida:2021aa}.\\ 
The main interface of OpenIPDM provides navigation features for accessing and obtaining the results of analyses on real data. Figure \ref{FIG:data_io} shows the hierarchical access to information starting from the structure-level to the output, which correspond to the deterioration state estimates for a structural element selected by the user.
From Figure \ref{FIG:data_io}, the information associated with each bridge $\mathcal{B}_j$ are dynamically stored in a \MATLAB{} array, which offers a better computational efficiency, relative to applying search queries on the entire database to access information. The computational efficiency can be further improved by pre-processing and storing the information of each bridge locally as \say{.mat} files. This is possible by accessing the function \emph{Pre-process Data} from the \emph{File} menu in OpenIPDM interface. In order to illustrate the input-output data in OpenIPDM, an example for a structural element $e_1^{130}$ is shown in Figure \ref{FIG:example_analyses}. In this figure, the deterioration model provide the forecast for the deterioration condition of the structural element $e_1^{130}$, based on the inspection data $\tilde{\bm{y}}_{t,1}^{130}\in[25,100]$. In this context, $\tilde{{y}}_{t,1}^{130}=100$, refers to a perfect condition for the structural element, while $\tilde{{y}}_{t,1}^{130}=25$, represent a poor condition for the structural element.
\\
\FloatBarrier
The main interface of OpenIPDM also provides access to additional toolboxes that enable essential functionalities in the software. Figure \ref{FIG:main_ui} shows a summery for the architecture of the software. 
\\
In Figure \ref{FIG:main_ui}, the software OpenIPDM takes as an input two types of file format, \say{.csv} which mainly contains the real database information, and \say{.mat} format which is employed for storing model parameters, pre-processed data and synthetic datasets. Moreover, the main interface of OpenIPDM has two important menus, namely the \emph{Tools} menu and the \emph{Analyses} menu, which are detailed in the next section.
\FloatBarrier
\subsection{Software functionalities}
This section provides an overview for the main functionalities in OpenIPDM, which are accessible through either the \emph{Tools} menus or the \emph{Analyses} menu.
\subsubsection{Software tools}
The functionalities in the \emph{Tools} menu are related to pre-processing real data, parameter estimation for the deterioration framework and generating synthetic data. The toolboxes available in the \emph{Tools} menu are briefly described in the following list:
\begin{itemize}
\item Read Real Database: the purpose of this toolbox is to pre-process the real visual inspections database, where it takes \say{.csv} files as an input and produces \say{.mat} files ready for analyses in \MATLAB{}. The extracted \say{.mat} files mainly contain information about the inspections and structural attributes, which are arranged in a \MATLAB{} cell-array variable following the template shown in Table \ref{cellarray_insp}.
\begin{table}[hbt]
\captionsetup{width=.75\textwidth}
\centering
\caption{Example for data structure of the \MATLAB{} cell-array variable containing information about the elements and structures.} 
\begin{tabular}{c |c c c|}
\cline{2-4}
& $e_1$ & $\dots$ & $e_\mathtt{E}$ \\ 
\cline{1-4}
\multicolumn{1}{ |c| }{structure \#1} & $[_{\mathtt{T}\times\mathtt{V}}]$ & $\dots$ & $[_{\mathtt{T}\times\mathtt{V}}]$      \\
\multicolumn{1}{ |c| }{$\vdots$} & $\vdots$ & $\vdots$ & $\vdots$      \\ 
\multicolumn{1}{ |c| }{structure $\#\mathtt{B}$} &  $[_{\mathtt{T}\times\mathtt{V}}]$ & $\dots$ & $[_{\mathtt{T}\times\mathtt{V}}]$      \\
 \cline{1-4}
\end{tabular}
\label{cellarray_insp}
\end{table}\\
The $\mathtt{B}$ rows in the cell-array correspond to the total number of structures that are visually inspected, while each column correspond to an element in the structure, with an array of input data $\mathtt{T}\times\mathtt{V}$. The $\mathtt{T}$ rows represent the total number of inspections performed on the structural element, while the $\mathtt{V}$ columns represent the element's information imported by the user. The information for each element include (from left-to-right): observations $\tilde{y}$, inspector ID $I_i$, inspection year $t$, structure ID, element's material identifier, structure's age and a subset of structural attributes selected by the user.

\item Generate Synthetic Database: generating synthetic data is important for verifying any hypothesis in the deterioration model, performing controlled experiments and assessing the performance of the state estimation as well as parameter estimation framework. The toolbox ensures that the generated dataset is quantitively and qualitatively representative of the real inspections database, through several criteria defined in the work of Hamida and Goulet \cite{Hamida:2020aa, Hamida:2020ab, Hamida:2021aa}.

\item Model Training: there exist three model training toolboxes, the first toolbox is for estimating the deterioration model parameters for a single structural category (e.g., $\mathcal{S}^j_1$: \{all the elements of type \emph{beam}\}), the second toolbox is for estimating the deterioration model parameters for multiple structural categories (e.g., $\mathcal{S}^j_{1:\mathtt{S}}$: \{beams, slabs, \dots, etc\}), while the third toolbox is for estimating the model parameters associated with the interventions framework. Prior to the training step, the user can determine the size of the training-validation-testing sets, which are employed to avoid overfitting. The parameter estimation functions rely mainly on the maximum likelihood estimate (MLE) method, and updates the parameters using a gradient-based optimization approach, and in some instances, the parameter estimation is done using online inference (e.g., estimating the inspectors parameters) \cite{Hamida:2020aa,Deka2021}. Due to the high number of structural elements in a normal case scenario (e.g., $\mathtt{E}\gg 10^3$), the majority of operations throughout the parameter estimation are vectorized and rely mainly on using (GPU) as well as parallel computing using (CPU). 
\end{itemize}
\subsubsection{Software analyses}
The \emph{Analyses} menu in OpenIPDM provides access to model validation and verification analyses using real and synthetic data, respectively. The toolboxes under this menu are:
\begin{itemize}
\item Model Verification: the verification analyses on synthetic data involves assessing the predictive capacity of the deterioration model by comparing the estimated deterioration state with the true state. There are different verification assessments, such as, examining the forecast error over time, and investigating any biases in the estimation. Furthermore, it is also possible to verify the estimation of the model parameters with the true parameters, especially, the parameters associated directly with the inspectors. 

\item Model Validation: this toolbox validates the deterioration model performance by relying on inspection data that is never employed in training the deterioration model. In order to perform the validation using this toolbox, it is required to have two complete real databases with one database containing more inspection data than the other. The additional inspection data is considered as a test set, which can be utilized by the toolbox to determine if any bias exist in the deterioration model forecast or to report the log-likelihood of the additional inspection data.
\end{itemize}

\subsubsection{Unit tests}
In order to verify and maintain the different functionalities in the software, a number of unit tests are created and embedded in the interface. The main purpose of the unit tests is to verify the deterioration model performance using different sets of input. The unit tests are available and accessible from the \emph{Developer} menu, for toolboxes with analyses and/or parameter estimation functionalities.

\section{Illustrative Example}
In this section, an example for generating and using synthetic data is presented. The example case include three steps : 1) generating synthetic data, 2) estimating model parameters, and 3) performing verification analyses. Each of the aforementioned steps are detailed in the following sections.

\subsection{Generate Synthetic Dataset}
In order to generate synthetic dataset, the toolbox \emph{Generate Synthetic Data} is employed, which is accessible from the menu \emph{Tools} in the main menu bar of OpenIPDM. The interface for this toolbox is simple, and requires mainly the model parameters as an input. For this example case, we generate inspection data for $\mathtt{E}=20000$ structural elements, with the average service-life for each structural element is $\mathtt{T} = 60$ years. The total number of inspectors $\mathcal{I}=\{I_{1},I_{2},\dots,I_{\mathtt{I}}\}$ involved is $\mathtt{I}=223$ synthetic inspectors. Each synthetic inspector is assumed to have an observation error model, $v_{t}:V\sim\mathcal{N}(\mu_{V}(I_{i}),\sigma_{V}^2(I_{i}))$. The standard deviation $\sigma_{V}(I_{i})$ and bias $\mu_{V}(I_{i})$ are generated for each synthetic inspector using a uniform distribution and according to the user defined values. In this example, we rely on the default input in the GUI, with minor adjustments shown in Table \ref{gendata_config}. 
\begin{table}[hbt]
\centering
\caption{Input required to configure and generate synthetic dataset using the \emph{Generate Synthetic Data} toolbox.}
\begin{tabular}{|llll}
\hline
\multicolumn{4}{ c }{ Configuration of Example Synthetic Dataset} \\ \hline\hline
\multicolumn{1}{l}{Time Span} & \multicolumn{1}{l|}{60} & \multicolumn{1}{l}{$\sigma_{V}(I_{i})$} & $[1, 6]$  \\
\multicolumn{1}{l}{Number of Time Series} & \multicolumn{1}{l|}{20000} & \multicolumn{1}{l}{$\mu_{V}(I_{i})$} & $[0,0]$  \\
\multicolumn{1}{l}{Number of Inspectors} & \multicolumn{1}{l|}{223} & \multicolumn{1}{l}{Trans. Param.} & 4 \\ \hline
\end{tabular}
\label{gendata_config}
\end{table}\\
After providing the input, it is possible to start the process of generating the synthetic dataset by pressing the button \emph{Generate}. The time to generate the synthetic data mainly depends on the size of the dataset and the compatibility of configurations with the monotonic deterioration process. The output of this toolbox is a set of \say{.mat} files, which is obtained after pressing the \emph{Save} button. The list of files is composed of,
 \begin{itemize}
\item Generated Data: contains all the information involved in generating the synthetic inspection dataset.
\item Full and Short: contains the true state and the corresponding sampled observations.
\item InspectorID: has the unique identifier (ID) for each synthetic inspector.
\item Observed: contains the synthetic inspections data. 
\item TrueInspector: includes the inspectors true $\sigma_V(I_i)$, bias $\mu_V(I_i)$ as well as the synthetic inspectors IDs, which collectively can be employed to verify the parameters estimation process.
\end{itemize}
For additional information about the data generation step, the user can refer to the manual available with the software.
\subsection{Model Parmaters Estimation}
In order to estimate the deterioration model parameters based on the generated synthetic inspection dataset, we use the toolbox \emph{Model Training - Generic}, accessible from the menu \emph{Tools} in the main menu bar. The input to this toolbox include loading the files: \say{Observed.mat} and \say{InspectorID.mat} in the interface, using the load button. Since, the dataset in this example is synthetic, the user has to select the checkbox \emph{Synthetic Data} on the interface. Thereafter, the user has to specify the deterioration model type and the optimization algorithm. In this example, we consider the use of the SSM deterioration model and rely on \emph{Newton-Raphson} for the parameter estimation \cite{Hamida:2020aa}.\\ 
The next step is to configure the parameters bounds and the test set size, which is done by pressing the button \emph{Model Config}. The \emph{Model Config.} interface allows defining the parameters bounds, as well as other options relating to the parameter estimation process. For this example case, the user can rely on the default configuration for the model parameters, and save the sittings by pressing the button \emph{OK}. After this step, the user can processed to estimate the deterioration model parameters according to the provided configurations. It should be noted that the parameters estimation step requires access to parallel computing toolbox in \MATLAB{} as well as a dedicated GPU card. At the end of the parameter estimation process, the toolbox provides two \say{.mat} files containing the model parameters and the inspectors parameters.
\subsection{Verification Analyses \& Results}
\begin{figure*}[hbt!]
\centering
\includegraphics[width=0.9\textwidth]{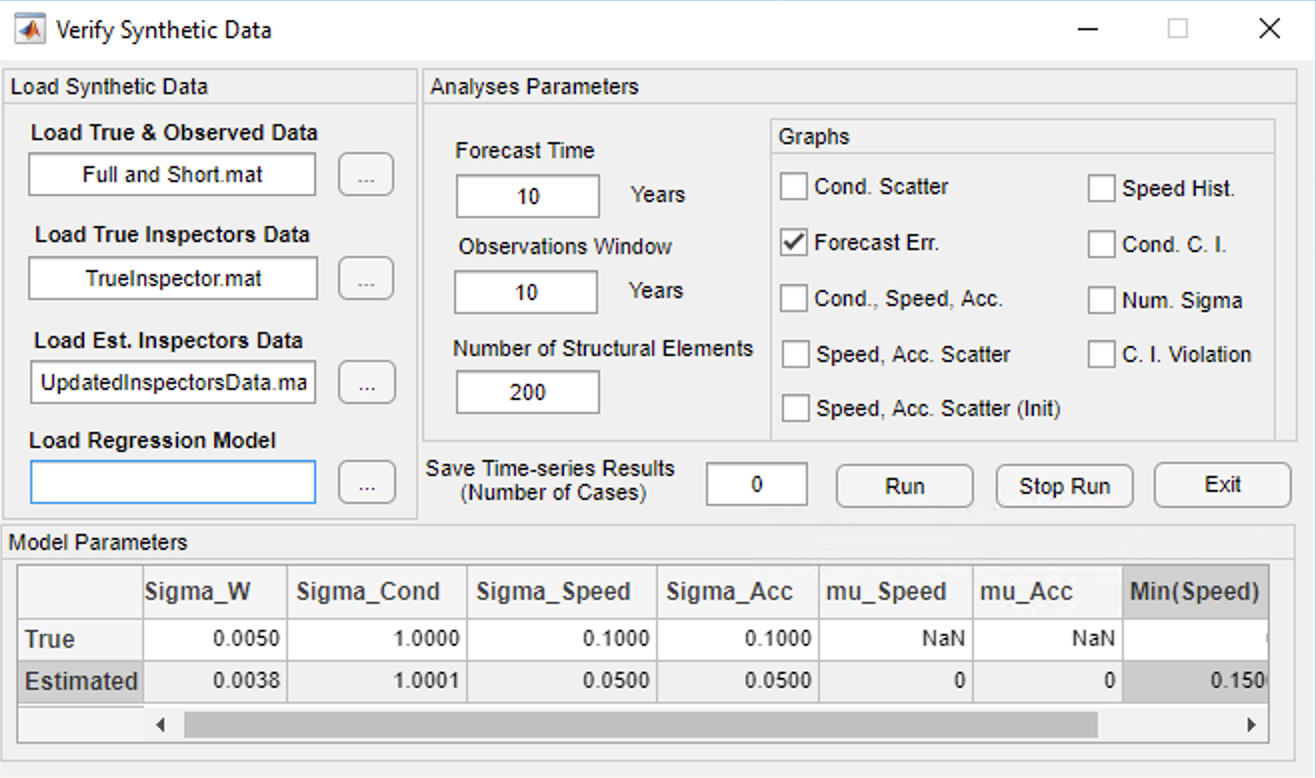}
\caption{Input and configuration for the toolbox \emph{Verify Synthetic Data}.}
\label{FIG:example_case}
\end{figure*}

\begin{figure*}[hbt!]
\centering
\subfloat{\includegraphics[width=0.3\textwidth]{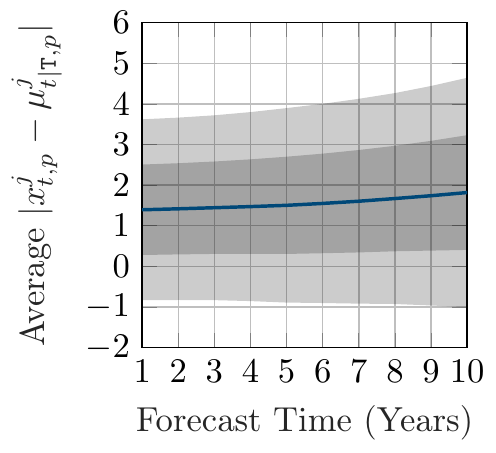}}
\subfloat{\includegraphics[width=0.3\textwidth]{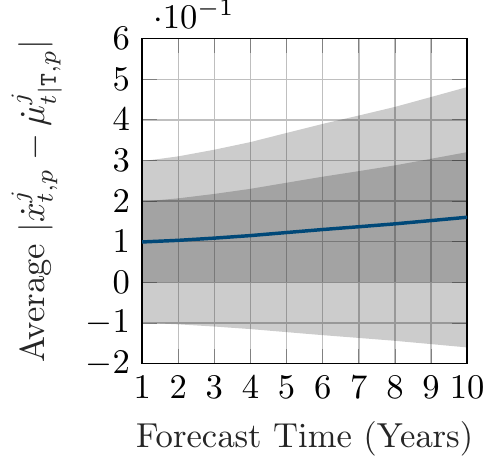}}
\subfloat{\includegraphics[width=0.3\textwidth]{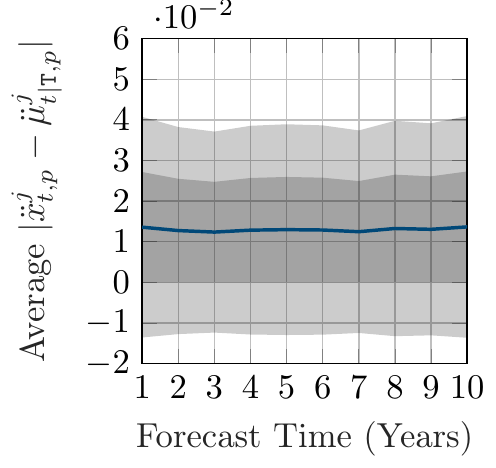}}
\caption{\label{FIG:muTrueVsEstimate}Absolute average error in forecast time for the expected condition, speed and acceleration based on the true condition, speed and acceleration respectively, with the 95\% confidence interval ($\pm2\sigma$) for each error.} 
\end{figure*}

\begin{figure*}[hbt!]
\centering
\subfloat{\includegraphics[width=0.3\textwidth]{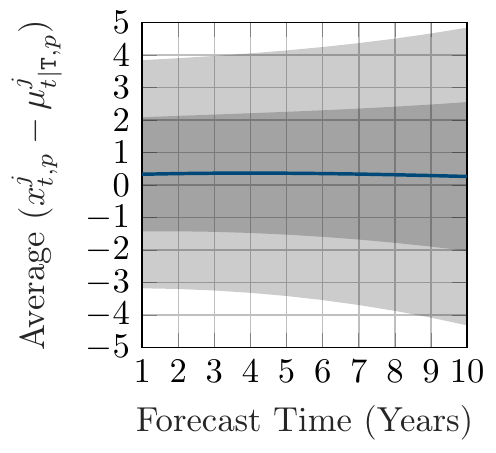}}
\subfloat{\includegraphics[width=0.3\textwidth]{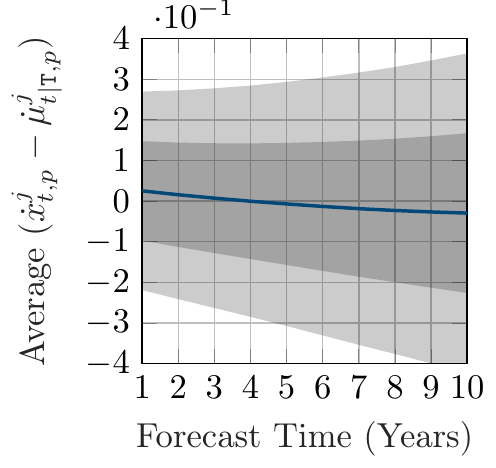}}
\subfloat{\includegraphics[width=0.3\textwidth]{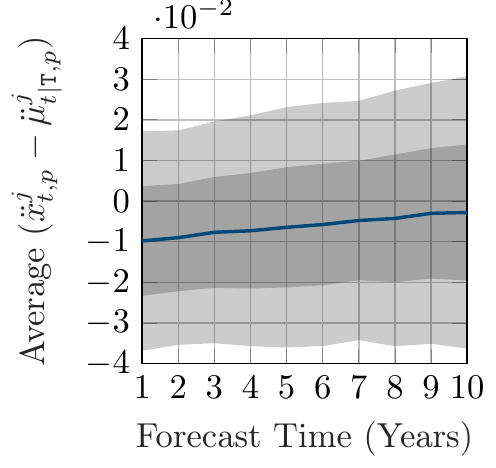}}
\caption{\label{FIG:muTrueVsEstimateAv} Average error in forecast time for the expected condition, speed and acceleration based on the true condition, speed and acceleration respectively, with the 95\% confidence interval ($\pm2\sigma$) for each error.} 
\end{figure*}

The last step in this case study is to examine the predictive capacity of the deterioration model using the model parameters estimated in the previous step. This is done by using the toolbox \emph{Verify Synthetic Data}, accessible from the \emph{Analyses} menu in the main menu bar of OpenIPDM. There exist several assessments that can be performed in order to verify the deterioration model performance. In this example, we consider the average forecast error over a period of time $\mathtt{T} = 10$ years, for a group of $\mathtt{E} = 200$ synthetic structural elements. The input to the toolbox is demonstrated on the interface shown in Figure \ref{FIG:example_case}. After providing the input files and parameters, it is possible to execute the verification analyses by pressing the button \emph{Run}.
\\
The processing time for the verification analyses mainly depends on the number of synthetic structural elements $\mathtt{E}$ involved in the analyses. After the analyses are done, the software output two graphs, shown in Figures \ref{FIG:muTrueVsEstimate} and \ref{FIG:muTrueVsEstimateAv}. The first graph shows the absolute average error for the model forecast of the expected deterioration condition, speed and acceleration, as shown in Figure \ref{FIG:muTrueVsEstimate}.
On the other hand, the second graph shows the average error for the model forecast of the expected deterioration condition, speed and acceleration (Figure \ref{FIG:muTrueVsEstimateAv}).
Figures \ref{FIG:muTrueVsEstimate} and \ref{FIG:muTrueVsEstimateAv} represent the outcome of a single assessment, however, there are other verification assessments, which are described in details within the software manual.
\FloatBarrier
\section{Impact}
OpenIPDM represents a step towards the accessibility to information about network-scale visual inspection data, as well as the deterioration behaviour of infrastructures. In many instances, visual inspection databases are managed by private and/or government agencies, therefore, the access to the inspection data is limited. OpenIPDM overcomes the data availability by enabling users to generate synthetic datasets with similar characteristics to visual inspection data taken from a real case study. Moreover, the software provides insights into  managing network-scale visual inspections and provides a template for pre-processing and organizing the information for an efficient computational time.\\
In addition, OpenIPDM relies on probabilistic methods with a demonstrated capacity to scale and a high potential for improvements in different areas. Promising improvement directions in this context are, enhancing the deterioration model by incorporating additional information about infrastructures, as well as reducing the computational cost associated with the parameter estimation framework. Furthermore, the software provides solid foundations for the development and testing of decision-making and planning frameworks.

\section{Conclusions}
In this paper, we present OpenIPDM, an open-source software for modelling the deterioration behaviour of infrastructures. The software is composed of several toolboxes which collectively enable the design, validation and verification of network-scale deterioration models. The frameworks currently available in OpenIPDM has state-of-the-art capabilities in terms of estimating the inspectors' uncertainties and modelling the effect of interventions based on visual inspections. In addition, the software provides tools for managing large visual inspections databases as well as handling missing data and outliers. OpenIPDM is open for contributions and users are encouraged to extend the software capacity in modelling the deterioration behaviour and/or develop additional functionalities such as, the development of decision making and planning modules.

\section*{Acknowledgements}
This project is funded by the Transportation Ministry of Quebec Province (MTQ), Canada. The authors would like to acknowledge the support of Simon Pedneault and Ren\'e Gagnon for facilitating the access to information related to this project.

{
\bibliographystyle{plainnat}
\bibliography{CitedWork_ZH}
}

\section*{Required Metadata}
\label{}

\section*{Current code version}
\label{}

\begin{table}[!h]
\begin{tabular}{|l|p{6.5cm}|p{6.5cm}|}
\hline
\textbf{Nr.} & \textbf{Code metadata description} & \textbf{Please fill in this column} \\
\hline
C1 & Current code version & v1.0.0 \\
\hline
C2 & Permanent link to code/repository used for this code version & \url{https://github.com/civml-polymtl/OpenIPDM/} \\
\hline
C3  & Permanent link to Reproducible Capsule & N.A.\\
\hline
C4 & Legal Code License   & MIT Licenses \\
\hline
C5 & Code versioning system used & git \\
\hline
C6 & Software code languages, tools, and services used & \MATLAB{} \\
\hline
C7 & Compilation requirements, operating environments \& dependencies & \MATLAB{} (version 2020b or higher), the \MATLAB{} statistics and machine learning toolbox, and access to GPU and parallel computing (required only for Model Training toolbox).\\
\hline
C8 & If available Link to developer documentation/manual & \url{https://github.com/CivML-PolyMtl/OpenIPDM/tree/main/Help} \\
\hline
C9 & Support email for questions & \url{zac.hamida@gmail.com}\\
\hline
\end{tabular}
\caption{Code metadata (mandatory)}
\label{} 
\end{table}

\end{document}